# Inverse Design of Block Polymer Materials with Desired Nanoscale Structure and Macroscale Properties

Vinson Liao[1], Arthi Jayaraman[1, 2, 3] *

1. Department of Chemical and Biomolecular Engineering, University of Delaware, Colburn Lab, 150 Academy Street, Newark, DE 19716
2. Department of Materials Science and Engineering, University of Delaware, 201 DuPont Hall, Newark, Delaware 19716,
3. Data Science Institute, University of Delaware, Ammon Pinizzotto Biopharmaceutical Innovation Center, Suite 147, 590 Avenue 1743, Newark, DE 19713

*Corresponding author: arthij@udel.edu

**Abstract**

The rational design of novel polymers with tailored material properties has been a long-standing challenge in the field due to the large number of possible polymer design variables. To accelerate this design process, there is a critical need to develop novel tools to aid in the inverse design process and to efficiently explore the high-dimensional polymer design space. Optimizing macroscale material properties for polymeric systems is even more challenging than inorganics and small molecules as properties are dictated by features on a multitude of length scales, ranging from the chosen monomer chemistries to the chain level design to larger-scale (nm to microns) domain structures. In this work, we present an efficient high-throughput in-silico based framework to effectively design high-performance polymers (blends, copolymers) with desired multi-scale nanostructure and macroscale properties, which we call RAPSIDY 2.0 - Rapid Analysis of Polymer Structure and Inverse Design strategY 2.0. This new version of RAPSIDY builds upon our previous work, RAPSIDY 1.0, which focused purely on identifying polymer designs that stabilized a desired nanoscale morphology. In RAPSIDY 2.0 we use a combination of molecular dynamics (MD) simulations and Bayesian optimization driven active learning to optimally query high-dimensional polymer design spaces and propose promising design candidates that simultaneously stabilize a selected nanoscale morphology and exhibit desired macroscale material properties (e.g., tensile strength, thermal conductivity). We utilize MD simulations with polymer chains preplaced into selected nanoscale morphologies and perform virtual experiments to determine the stability of the chosen polymer design within the target morphology and calculate the desired macroscale material properties. Our methodology directly addresses the unique challenge associated with copolymers, whose macroscale properties are a function of both their chain design and mesoscale morphology, which are coupled. We showcase the efficacy of our methodology in engineering high-performance blends of block copolymers that exhibit (1) high thermal conductivity and (2) high tensile strength. We also discuss the impact of our work in accelerating the design of novel polymeric materials for targeted applications.

## I. Introduction

The rational design of novel materials with targeted properties and functionalities has been a longstanding scientific challenge in fields working on drug delivery[1], renewable energy[2], and microelectronics[3]. Material design has traditionally relied on the forward design process, a trial-and-error based approach where candidates are systematically proposed, synthesized, and tested to determine if the material property satisfies the user-defined objective[4]. While the forward design process is conceptually straightforward, this empirical methodology is both time-consuming and resource-intensive due to the vast size of the chemical space. The design problem becomes even more challenging for macromolecular materials (i.e., polymers), as opposed to small organic molecules (which already pose significant difficulty) or inorganic materials, as the number of potential designs combinatorically explodes due to the large number of independent parameters (e.g., number of monomer units, polymer molecular weight, branching patterns, composition, and sequence arrangements).

Within macromolecular materials, block copolymers (BCPs) have been of great interest to materials researchers due to their ability to self-assembly into a variety of ordered nanostructures, or morphologies (e.g., lamellae, hexagonally packed cylinders, and double gyroids) with some of these morphologies being uniquely suited for specific applications. For example, the double gyroid morphology is ideal for applications that desire good mass or charge transport, as in membranes[5] and photovoltaics[6], respectively. The formation of the various nanostructures is driven by balancing of enthalpic forces (which favor de-mixing due to chemical incompatibility) between monomer chemistries and entropic forces (which favor mixed states)[7–9]. This interplay between enthalpic and entropic forces is highly sensitive to design parameters such as monomer volume fraction, chain length, block length, and sequence architecture.

BCPs' large design parameter space can easily be illustrated by considering the design space even for the simplest BCP, the linear diblock copolymer with two chemistries, A and B. The design parameters that one could independently vary to tune the nanoscale ordered morphology are: (1) the chosen monomer chemistries, (2) the monomer segregation strength (or temperature), (3) the block compositions, and (4) the total degree-of-polymerization (DP). An additional layer of difficulty arises due to the fact that macroscale BCP properties are governed by attributes on a multitude of length scales, ranging from the molecular level (i.e., the monomer chemistries) to the chain level (i.e., block composition and DP) to the domain level (i.e., the morphology formed), all of which are coupled in ways that are not known a priori. Thus, the number of unique BCP designs that make up the parameter space for even the simplest case of BCPs - the linear diblock - is intractably large. As advances in synthetic techniques push the boundaries of the chemical (number of monomer chemistries) and physical heterogeneities (branching, dispersity) that can be precisely programmed into polymers, the corresponding design parameter space becomes unfeasible to be explored using naïve forward design methodologies to satisfy the user-defined objective[10–14]. As a result, the rational design of polymers, BCPs in particular, with tailored macroscopic properties has been a longstanding challenge and has motivated the need for new computational methodologies and high-throughput automated experiments to develop accurate structure-property relationships and accelerate material design.

One of the ways that scientists have attempted to accelerate material design is by reversing the forward design process, a process known as inverse design. Rather than beginning with a set of proposed design parameters and experimentally determining the resultant material property vis-à-vis the forward design process, in the inverse design process, one begins with a target objective and directly searches for designs that meet the user-defined criterion (criteria)[13]. Intelligent and efficient navigation of high-dimensional design spaces is required to address the intrinsic challenge that plagues inverse problems; they often are ill-posed (or weakly conditioned) and may not have unique solutions. Novel computational and machine learning based tools have been the foundation for tackling these unique challenges, especially within the high-performance polymer design space.

For example, Vogel and Weber utilized generative models known as variational autoencoders (VAE) to encode high-dimensional polymer structures (including stoichiometry and chain architecture) into continuous, low-dimensional latent spaces that enabled high-throughput search[15]. They utilized global optimization methods such as Bayesian optimization (BO) and genetic algorithms (GA) to traverse the latent space to target polymers with superior electronic properties for photocatalytic applications. Work by Zhou *et al.* utilized GAs coupled with atomistic molecular dynamics simulations to optimize polyethylene-polypropylene (PE-PP) sequences for high thermal conductivity[16]. Using GAs, they were able to discover "high-performance" polymer sequence designs that exceeded the thermal conductivity of the parent PE and PP homopolymers. Work by Wang *et al.* utilized BO with coarse-grained (CG) molecular dynamics (MD) simulations to identify optimal CG polymer design parameters (such as molecule size and intermolecular interaction strength) to design highly conductive polymer electrolytes. By iteratively searching using BO, they were able to identify CG parameters that led to an order-of-magnitude increase in ionic conductivity compared to brute-force forward design search in the same number of simulations. While computational and machine learning based methods have found success in designing high-performance polymeric materials in general, the same success cannot be said for designing high-performance BCPs. Unlike, for instance, homopolymers, random copolymers, or statistical copolymers, macroscale material properties of BCPs are a function of both their sequence design and mesoscale morphology, which are coupled. Thus, current inverse design methodologies of BCPs have primarily focused on exploring the high-dimensional sequence design space to identify viable sequence designs for desired morphologies, rather than macroscopic properties.

Most BCP inverse design-based methodologies rely on the well-established field-theoretic methods as a means of forward design in which discrete polymer chains are approximated in a mean-field manner and obey continuous governing equations. Energetic interactions are described using an effective Hamiltonian operator, $\mathcal{H}$, that is dependent of system parameters such as effective interaction strength and chain architecture. Field-theoretic methods, and self-consistent field theory (SCFT) in particular, have shown success in computing experimentally determined phase diagrams for BCPs, predicting order-disorder transition (ODT) temperatures, and discovering novel BCP morphologies[17–19]. Recent literature has found success in combing field-theoretic methods with machine learning to optimize BCP designs for targeted morphologies.

For example, Tsai and Fredrickson utilize SCFT in conjunction with a global optimization algorithm called particle swarm optimization (PSO) to develop a software known as PSO-SCFT to determine novel globally and meta-stable morphologies for BCPs[20]. PSO-SCFT can reproduce and recover globally stable morphologies of well-studied systems, such as AB diblock copolymers and mikoarm star polymers $AB_4$ and even discovered a novel energetically competitive mystery morphology for the latter system. Work by Khadilkar *et al.* also utilize PSO in conjunction with SCFT but developed a new methodology to determine and target polymer sequences that exhibited difficult to synthesize 3D network morphologies within the high-dimensional parameter space[21]. They used PSO and SCFT to successfully optimize design parameters such as species volume fraction and monomer interaction strength directly to target user-defined morphologies. While field-theoretic methods combined with machine learning have shown success in accelerating BCP design, mean-field methods (1) sacrifice molecular level details (e.g., chain configurations), (2) are only valid at the limit of infinitely long chains, and (3) neglect excluded volume interactions. In addition, field-theoretic methods lack the ability to drive optimization of macroscopic material properties, such as transport and mechanical properties[22,23], unlike more expensive particle-based simulations like molecular dynamics (MD) simulations.

In this work, we introduce a high-throughput in-silico framework for the inverse design of high-performance self-assembled block copolymers using a combination of MD simulations and BO driven active learning. Our approach is called RAPSIDY Rapid Analysis of Polymer Structure and Inverse Design strategY 2.0. This version of RAPSIDY builds upon our earlier version, RAPSIDY 1.0, which was a MD based methodology that significantly accelerated the stability analysis of BCP chain designs in target morphologies[24]. RAPSIDY 1.0 utilized MD simulations and an external guiding potential to preplace coarse-grained chains of a chosen polymer sequence design within a specified target morphology. This allowed us to directly assess the stability/meta-stability of a chosen polymer design within a target morphology without waiting for the self-assembly process from a disordered state, which is often slow due to sluggish system dynamics in dense polymer melts[25]. We showed that RAPSIDY 1.0 accelerates design space exploration by two orders-of-magnitude over traditional MD simulations in our example case study with simple naïve grid search. RAPSIDY 2.0 now builds upon its previous generation to optimize the BCP coupled parameters of chain designs and morphology for *high-performance macroscopic properties*, such as thermal conductivity and tensile strength. Our methodology utilizes a molecular dynamics driven process we call *virtual experiments* to determine (1) the stability of a chosen chain design with a target morphology and (2) the material's macroscopic property. We utilize Bayesian Optimization driven active learning to "learn" from our *virtual experiments* to optimally navigate the design space and propose promising candidates for follow up virtual experimentation. We showcase our methodology on designing blends of linear diblock BCPs with high thermal conductivity and high tensile strength. Our framework offers an efficient and robust route to discover novel, high-performance BCP designs with tailored macroscopic materials and helps to overcome the longstanding challenge of multiscale design coupling.

## II. RAPSIDY 2.0
### A. Overview

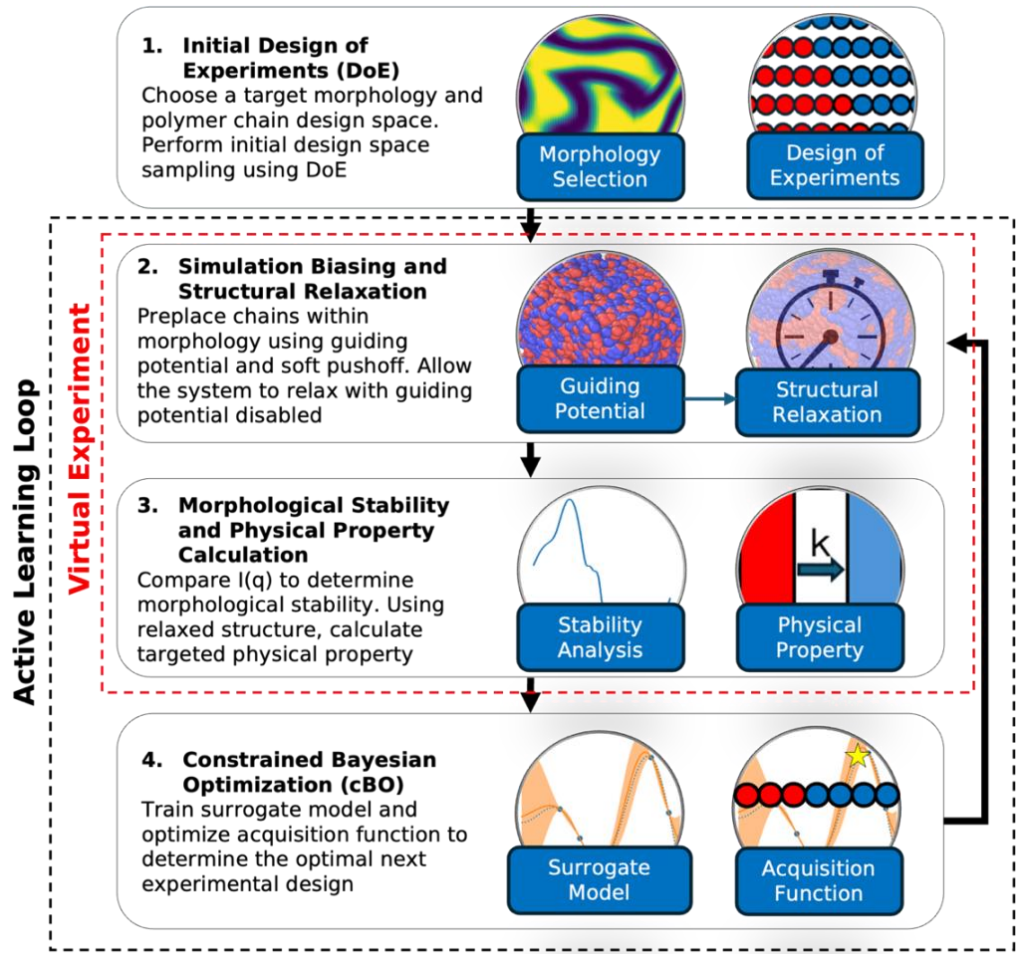

*Figure 1.* **Overview of RAPSIDY 2.0 framework.** Our methodology begins by selecting the targeted morphology and mathematically defining the targeted polymer design space. We perform initial sampling using standard design of experiments (DoE) techniques such as Latin hypercube sampling (LHS), full factorial design, and completely randomized design. Here, we use LHS due to its near-random design coupled with efficient hypervolume filling capabilities. Using the selected set of design parameters, we perform in-silico virtual experiments (simulations) to determine the morphological stability and the targeted physical property. We utilize our previous methodology, RAPSIDY 1.0, and preplace chains within the targeted morphology using a guiding potential and subsequently allow the system to relax with the guiding potential to determine compatibility between chain design and morphology. Unlike RAPSIDY 1.0, we do not screen multiple lattice constants (box sizes) in parallel and instead allow the active learning loop to determine optimal box sizes. Finally, we utilize constrained Bayesian optimization to determine the next optimal design to test by training cheap-to-evaluate surrogate models to predict both the morphological stability and the targeted physical property and optimize the acquisition function to determine the optimal design to perform virtual experiments on next to close the loop.

**Figure 1** shows a schematic of the RAPSIDY 2.0 framework, which is composed of three major steps: (1) *initial design of experiments*, (2) *virtual experiments*, and (3) *active learning loop*. The workflow begins with the *initial design of experiments* by mathematically defining the desired N-dimensional design space and desired target morphology. Using design of experiments (DoE) methodologies as described in various statistics textbooks[26], we sample the initial design space to generate a representative subset of data that will be later used to train our machine learning model and guide our active learning loop.

Next, for each of the sampled points within the design space, we perform *virtual experiments* to (1) determine the compatibility between the chosen design sequence and targeted morphology and (2) to calculate the material property using molecular dynamics. Our virtual experiments build upon our previously published methodology, RAPSIDY 1.0, which rapidly screens the stability and compatibility of copolymer(s) chain designs with target morphologies. Rather than allowing the system to naturally self-assemble into its free energy minima structure, we utilize a process that we call structural biasing to directly place the polymer chains into the simulation box within the targeted morphology using an external guiding potential and soft push-off potential. Rather than screening multiple lattice constants in parallel using a grid-search approach to determine compatible periodic simulation box sizes like in RAPSIDY 1.0, we treat the lattice constant as a design parameter (which is unique to in-silico studies of BCPs whose self-assembled morphology has a characteristic lattice constant) to be intelligently queried using machine learning. We then disable the guiding and push-off potentials and allow the system to naturally relax with its intrinsic bonded and non-bonded potentials to determine if the targeted morphology is a local minimum in the potential energy surface. We quantitatively determine the compatibility between the target morphology and polymer design (henceforth referred to as 'morphological stability') by comparing the computed scattering profiles after relaxation with the scattering profile of the biased structure. We now add an additional step by taking the relaxed structure and computing physical properties directly using (non-equilibrium) molecular dynamics to complete the *virtual experiment*.

We then begin the *active learning loop* by utilizing the data obtained from the first round of virtual experiments of the initial DoE samples to train surrogate models to predict the morphological stability and our physical property of interest, $f(\vec{x})$. Using a machine learning methodology known as *constrained Bayesian Optimization*, we utilize this surrogate model to determine the next experimental design to test that both maximizes our target property of interest and chooses designs that are compatible with the target morphology. The chosen design point is fed back into step (2) *virtual experiment* to determine its physical property and morphological stability, which is then used as additional training data in the *active learning loop*. The design loop is then closed and repeated until the desired property of interest is maximized with respect to the given user tolerance. In the subsequent sections, we elaborate on each step of RAPSIDY 2.0 and showcase the efficacy of our methodology in designing high-performance macromolecular materials composed of BCPs that exhibit high thermal transport properties and determine the effect of morphology on thermal conductivity. BCPs have emerged as a promising class of materials for applications such as microelectronics and semiconductors because of properties such as high ionic conductivity and convenient microdomain tunability but is limited by its ability to dissipate heat effectively[27–29]. Our case study of RAPSIDY 2.0 in subsequent sections seeks to address this limitation of BCPs.

For this work, we focus our efforts on studying a blend of linear diblock copolymers (diBCPs), modeled as $A_{x1}B_{y1}$ and $A_{x2}B_{y2}$, where x1, y1, x2, and y2 refer to the degree of polymerization each block. Even though we restrict our demonstration of our methodology in this paper to diBCPs, the methodologies developed in this work can be extended to any one of the intractable designs among the vast BCP design space (e.g., AB multiblock copolymers, ABC multiblock copolymers, blends of homopolymers and copolymers, varying architectures of BCPs).

### B. Polymer Coarse-Grained (CG) Model

We model each diBCP as a coarse-grained (CG) bead-spring chain with two types of beads (hereon, referred to as A and B) with each bead size chosen to be the statistical segment length of that monomer. We consider the case where the statistical segment length of beads A and B are equivalently set to $1\sigma$. Adjacent beads in each chain are connected via harmonic bonds with a potential of the following form: $U_{bond}(r) = k_{bond}(r - r_0)^2$. Here, r refers to the bead-bead separation measured from the center of each bead and $r_0$ is the equilibrium bond length, which is set to the arithmetic mean of the statistical segment length of the two species involved. The force constant is set to $k_{bond} = 50 k_b T/\sigma^2$. Each polymer chain in our study consists of $N_1$ or $N_2$ beads, respectively. All pairs of nonbonded beads interact using a cut-and-shifted LJ potential, $U_{cut}$, of the following form:

$$U_{cut}(r) = \begin{cases} 4\varepsilon_{ij}\left[\left(\left(\frac{\sigma}{r}\right)^{12} - \left(\frac{\sigma}{r}\right)^{6}\right) - \left(\left(\frac{\sigma}{r_{cut}}\right)^{12} - \left(\frac{\sigma}{r_{cut}}\right)^{6}\right)\right], & r < r_{cut} \\ 0, & r \geq r_{cut} \end{cases} \quad (1)$$

The cutoff radius, $r_{cut}$, is set to $2\sigma$. The depth of the energetic well, $\varepsilon_{ij}$, for self-interactions (A-A, B-B) and cross-interactions (A-B) between bead types i and j are related via the Flory-Huggin $\chi$ parameter and are defined as:

$$\chi = \frac{z}{k_b T}\left[\left(\frac{\varepsilon_{AA} + \varepsilon_{BB}}{2}\right) - \varepsilon_{AB}\right] \quad (2)$$

where $\varepsilon_{AA}$ and $\varepsilon_{BB}$ refer to strengths of self-interaction energies, $\varepsilon_{AB}$ refers to the strength of the cross-interaction, and z is the coordination number. We set the coordination number to z = 6 in accordance with previous literature[30,31]. In this study, we fix $\varepsilon_{AA} = \varepsilon_{BB} = 1 k_b T$ and vary $\varepsilon_{AB}$ appropriately for the targeted $\chi$ value. We choose a fixed design with $\chi = 1.2$ to simulate a fixed monomer chemistry pair with $\chi N$ of at least 60 for the chain lengths studied. In principle, this value of $\chi$ could also be a parameter in the search space of this work. We keep the $\chi$ fixed to mimic the scenario where the researcher is restricted to using a specific set of A and B monomers and is looking to find the optimal designs of blends of A and B containing copolymers.

### C. Initial Design of Experiments

In the *initial design of experiments*, we start by defining the N-dimensional design space and utilize expert knowledge to define the bounds of the optimization study. To showcase RAPSIDY 2.0, we will utilize a benchmark 6-dimensional design space of diblock copolymer blends to determine optimal chain designs and blending ratios to maximize the material's thermal conductivity with a lamellae target morphology. **Figure 2a** is a schematic that illustrates the design parameters and their corresponding upper and lower bounds that thermal conductivity will be optimized over. Five of the parameters ($N_1$, $N_2$, $f_{A,1}$, $f_{A,2}$, and $\varphi_1$) correspond to chain design and blend parameters which can be tuned experimentally.

However, the sixth and final design parameter, L, corresponds to the system lattice constant and governs the simulation box size; this parameter is unique to BCP simulations and references the well-known problem of "box size incommensurability" of in-silico experiments. Box size incommensurability is a challenge in simulating BCP self-assembly that occurs when there is incompatibility between the simulation box size and the intrinsic periodic unit cell of the system (which is not known a priori), which in turn leads to difficulty in the formation of specific morphologies[32,33]. One and two-dimensionally periodic morphologies such as lamellae and hexagonally packed cylinders, respectively, are more resilient to box size incommensurability as they can morphologically rotate within the 3D simulation box to minimize internal strain. However, three-dimensionally periodic phases, such as double gyroid, cannot morphologically rotate within the 3D simulation box and are much more difficult to form without a correct initial guess of box size. In our previous work, RAPSIDY 1.0 utilized a parallelized grid-search approach to screen multiple lattice constants to determine the optimal box size for the chosen design. However, we now treat the system lattice constant as a design parameter to be guided by our active learning methodology to reduce the number of virtual experiments needed.

Once the design space and corresponding bounds are defined, we utilize statistical design of experiments methodologies to sample the initial design space and generate a representative subset of data that will be later used to train our machine learning model and guide our active learning loop. There are various methodologies of sampling high-dimensional spaces effectively, such as full factorial design, fractional factorial design, central composite design, and Latin hypercube sampling (LHS). We use LHS due to its near-random design and efficient space filling capabilities. We initially sample ~100 points using LHS, which follows the conservative 10-20 times the dimensions for the initial number of samples as outlined in literature[34,35]. **Figure 2b** shows the pairwise relationships for the initial DoE sampling using LHS with 100 samples. The diagonal plots show the univariate distributions and demonstrate that each individual parameter is uniformly sampled throughout the 100 samples. The off-diagonal plots show pairwise correlations between variables and show that the samples are uncorrelated.

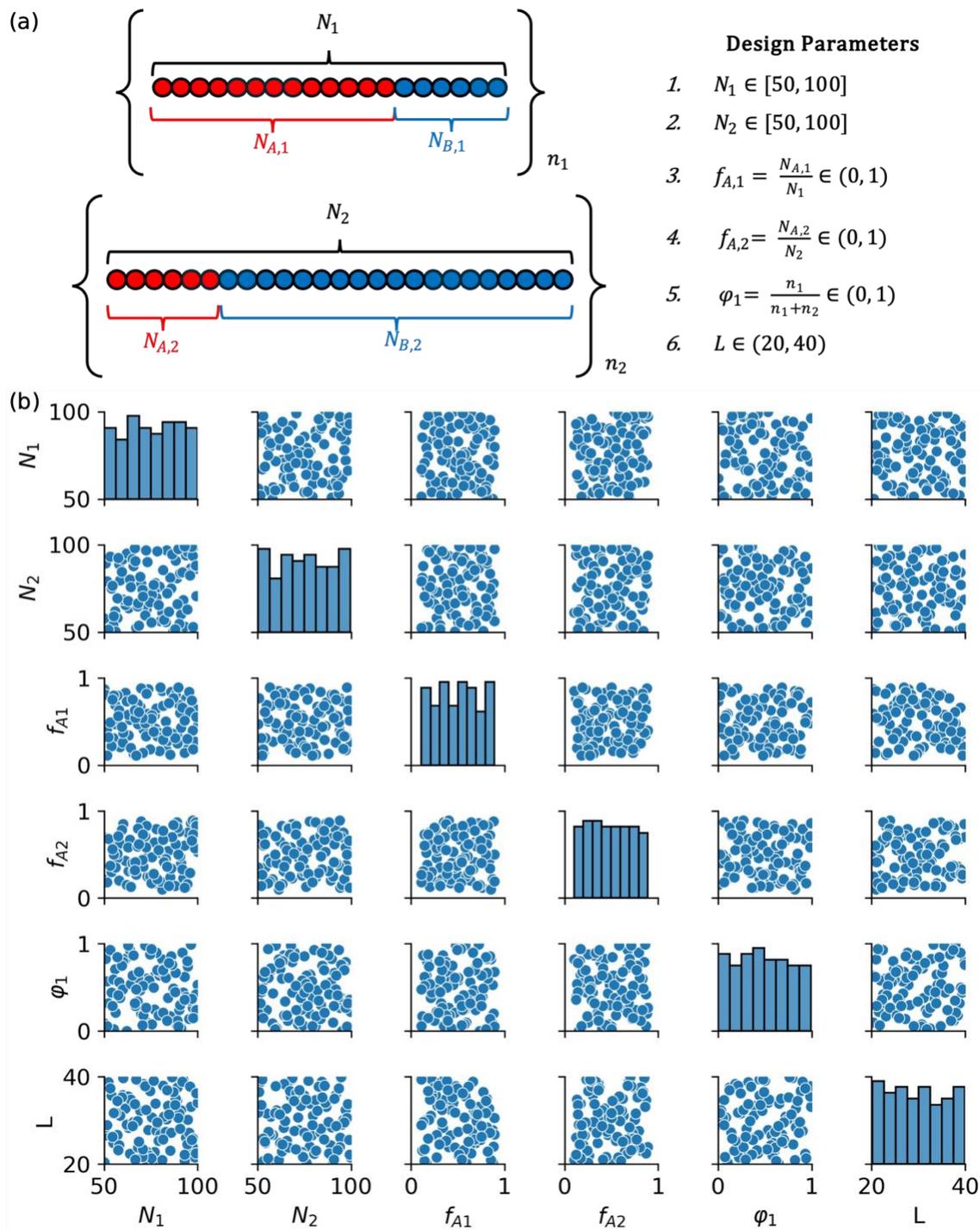

*Figure 2.* **Design Space and Initial Design of Experiments (DoE) for Diblock Copolymer Blends.** (a) The design space for blends of diblock copolymers is 6-dimensional, with five parameters related to chain and blend design parameters including chain length, volume fraction of A, and blend composition, and one parameter related to the simulation box size. The inclusion of the simulation box size parameter is unique for in-silico experiments of BCPs to address the well-known 'box size incommensurability' problem in standard BCP simulations, as described in the text. (b) Pairwise relationships for design of experiment sampling using

Latin hypercube sampling with 100 samples. Diagonal plots show univariate distributions and off diagonal plots show pairwise correlations.

### D. Running a Virtual Experiment

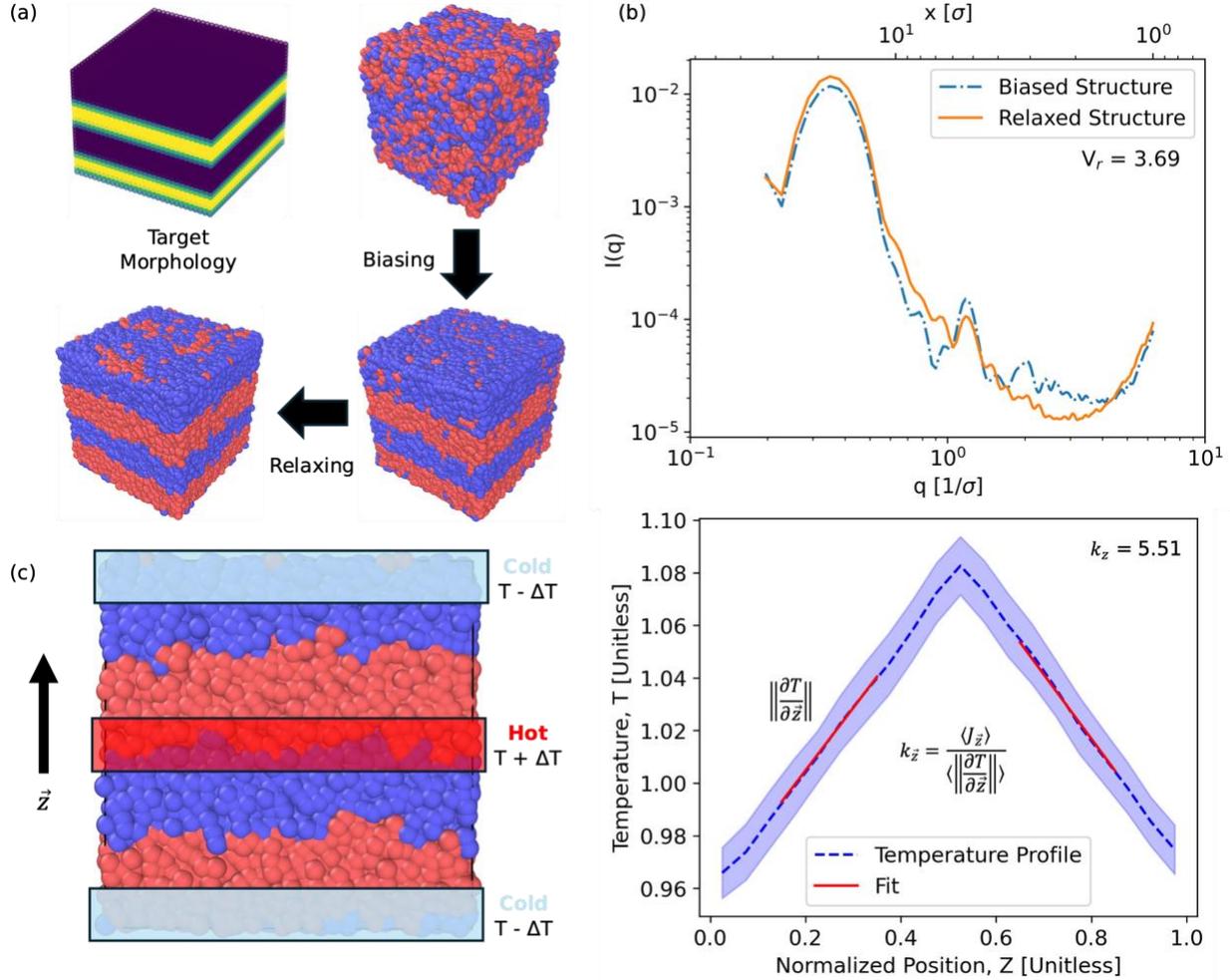

Figure 3. **Schematic showing steps of a virtual experiment for calculating the thermal conductivity using an example lamellar target structure.** (a) As done in RAPSIDY 1.0, a random disordered melt is biased into a target morphology (in this figure it is the lamellar phase). After the target morphology is achieved, the guiding potential is disabled to allow the system to naturally relax to its local energy minimum. (b) The compatibility of the target morphology with the chosen chain design is quantified using the volatility-of-ratio $V_r$ similarity between biased and relaxed structure scattering profiles. A lower value of $V_r$ implies higher stability. (c) Thermal conductivity calculations are performed using non-equilibrium molecular dynamics. The temperature profile is fit using a linear curve to determine the average temperature gradient and the value of thermal conductivity, $k_z$.

After completing the initial DoE and obtaining a representative sample of the design space, we perform virtual experiments to calculate the stability of the chosen designs within the targeted morphology as well as the corresponding physical property, which in our first case study is thermal conductivity in the z-direction. **Figure 3** shows a schematic of the process of running a virtual experiment for an example vector of design parameters $(N_1, N_2, f_{A1}, f_{A2}, \varphi_1, L) = (91, 58, 0.56, 0.61, 0.51, 39.58)$ with a target morphology of lamellae. We simulate $n_1$ chains of length $N_1$ and $n_2$ chains of length $N_2$ in a simulation box sized to match the sampled lattice constant such that a volume density of $\rho = 0.45$ is achieved to mimic dense melt conditions with our CG polymer model. Note that we may not be able to achieve the exact chain design parameters sampled because we are working with discrete chains in a finite size simulation box; thus, we choose discrete design values that best match the sampled parameter values.

At the beginning of our workflow, $M = (n_1 + n_2)$ chains are placed randomly in a cubic simulation box of the sampled box size. We then apply an external biasing potential of the following form to our random initial melt to generate configurations that match the target morphology exactly like our approach in RAPSIDY 1.0:

$$V_{ext} = A \sum_{i=1}^{M} [(\varphi_i^A - \varphi_i^{A,ref})^2 + (\varphi_i^B - \varphi_i^{B,ref})^2] \qquad (3)$$

Here, $\varphi_i^j$ and $\varphi_i^{j,ref}$ refer to the number density of bead type j within the simulation and target morphology, respectively, at mesh point i. M refers to the total number of mesh points and A is a positive user-defined scaling constant that controls the strength of the biasing potential (which we set to $A = 100$ to balance biasing strength with numeral stability). When the system is far away, the potential energy is heavily penalized but as the simulation approaches the target morphology the potential energy penalty tends to the limit of 0. Here, the density refers to the number density of each species (in this case, A or B) at a particular mesh point within the simulation box. We also utilize a soft push-off procedure[25], which slowly introduces excluded volume interactions to allow for accelerated equilibration dynamics, by modifying the bonded interactions to a force-capped LJ potential of the form:

$$U_{fc}(r) = \begin{cases} U_{cut}(r_{fc}), & r < r_{fc} \\ U_{cut}(r), & r \geq r_{fc} \end{cases} \qquad (4)$$

The cutoff radius is set to $r_{fc} = 0.8\sigma$ and increased linearly over 5000 timesteps $r_{fc} = 2^{1/2}\sigma$ over the course of the biasing process. We then perform molecular dynamics simulations within the canonical ensemble using the Large-scale Atomic/Molecular Massively Parallel Simulator[36] (LAMMPS) and evolve the system using the Nosé–Hoover thermostat at a reduced temperature of T*=1 for 5000 timesteps, which we found was the sufficient number of timesteps needed for all chains to settle within their targeted morphologies. We choose a timestep of $\Delta t = 0.01\tau$, where $\tau$ is in reduced LJ units of time.

After the simulation has reached its targeted morphology and achieved its "biased structure", we disable the external guiding potential and revert the soft push-off potential to the original hard non-bonded potential to allow the system to naturally evolve in the canonical ensemble to a local free energy minimum. We find that allowing the system to relax for $10^6$ timesteps is sufficient to determine if the target morphology is a local free energy minimum in the potential energy surface of our chosen chain design. The final structure after $10^6$ timesteps is referred to as the "biased structure". This biasing and relaxing procedure is visually shown in **Figure 3a** for our example $(N_1, N_2, f_{A1}, f_{A2}, \varphi_1, L) = (91, 58, 0.56, 0.61, 0.51, 39.58)$ with a target morphology of lamellae. Visually, we see that our example chain design is stable within the lamellae morphology after $10^6$ timesteps. To automate the stability analysis, we quantify the stability of the selected chain design in the target morphology using computed structure factors. We compare the computed structure factor (using the Debye equation[37]) of the biased structure and relaxed structure, as shown in **Figure 3b**, and utilize the volatility-of-ratio, $V_r$, as a distance metric to quantify the similarity[38]:

$$V_r = \sum_{i=1}^{25} \left| \frac{R(q_i) - R(q_{i+1})}{(R(q_i) - R(q_{i+1}))/2} \right| \tag{5}$$

$$R(q) = \frac{I_1(q)}{I_2(q)} \tag{6}$$

Scattering intensities, $I(q)$, are discretized over 25 equal size bins between $q_1 = \frac{2\pi}{D}$ and $q_{25} = \frac{2\pi}{d}$, where D and d refer to the shortest dimension of the simulation box and diameter of the smallest bead, respectively. The volatility-of-ratio provides a robust similarity metric to quantify structural similarity, where lower values of $V_r$ values imply a higher similarity between scattering profiles. We utilize $V_r$ and a predetermined cutoff using expert knowledge to compare I(q) before and after relaxation to determine compatibility between the chosen design and target morphology. In this work, we chose a cutoff of $V_r < 4.5$ to denote morphological stability. The example shown in

**Figure 3b**, $(N_1, N_2, f_{A1}, f_{A2}, \varphi_1, L) = (91, 58, 0.56, 0.61, 0.51, 39.58)$ with a target morphology of lamellae, has a similarity score of $V_r=3.69$, which is below our threshold value; by our cutoff criterion this means this selected set of design parameters are stable in this target lamellar morphology, consistent with visual observation.

The final step of the virtual experiment is to perform physical property calculations with the relaxed structure using molecular dynamics. In this example, we choose to perform thermal conductivity calculations using non-equilibrium molecular dynamics (NEMD) following the methodology outlined by Müller-Plathe[23]. In **Figure 3c** we show a schematic of the NEMD methodology for calculating thermal conductivity. An artificial temperature gradient is induced by exchanging the velocities between two particles (one of which is in the "hot" zone and one of which is in the "cold zone") in the simulation box every X timesteps, where X is a user-defined parameter. Because the exchanged velocities are known, the kinetic energy exchanged and the resulting heat flux can be easily computed. The temperature profile and temperature gradient are observed over time to determine the thermal conductivity in the direction of the imposed heat flux, which is given as:

$$k_{\vec{v}} = \frac{\langle J_{\vec{v}} \rangle}{\langle \left\| \frac{\partial T}{\partial \vec{v}} \right\| \rangle} \tag{7}$$

Here, $J_{\vec{v}}$ refers to the imposed heat flux, $dT/d\vec{v}$ refers to the observed temperature gradient, and $\vec{v}$ is the direction of heat flux. We perform kinetic energy swaps every 10 timesteps and observe the temperature profile over $10^6$ timesteps. We fit linear curves to the normalized positions regimes of [0.15, 0.40] and [0.60, 0.85] in the direction of $\vec{v}$ to determine the temperature gradient, as shown in Figure 3c. For the remainder of this work, we focus on thermal conductivity calculations in the z-direction.

The example shown in **Figure 3** with parameters $(N_1, N_2, f_{A1}, f_{A2}, \varphi_1, L) = (91, 58, 0.56, 0.61, 0.51, 39.58)$ and a target morphology of lamellae is found to have a z-directional thermal conductivity, $k_z$, of 5.51. To estimate the thermal conductivity in real units for an experimental system of interest, one needs to specify the following three paraments for the chosen polymer of interest (1) length scale ($\sigma$), (2) mass (m), and (3) interaction strengths ($\varepsilon$) in SI units to appropriately rescale the non-dimensional CG units[39]. The linear rescaling constant, $k_0$, is then given by:

$$k_0 = \frac{k_B \sqrt{\frac{\epsilon}{m}}}{\sigma^2} \qquad (8)$$

Here, $k_B$ refers to Boltzmann's constant. For instance, consider the following example of polystyrene, where each CG bead maps to a one styrene monomer, with the following parameters from literature which were derived to match experimentally observed density values[40]: $\sigma = 0.60*10^{-9}$ m, m = $1.73*10^{-25}$ kg, $\varepsilon = 5.31*10^{-21}$ J. The computed scaling constant is then $k_0$=0.0067 W/(m*K), which means our example design $(N_1, N_2, f_{A1}, f_{A2}, \varphi_1, L) = (91, 58, 0.56, 0.61, 0.51, 39.58)$ has a real unit thermal conductivity value of 0.037 W/(m*K) and is in good agreement with experimental values of 0.032 to 0.038 W/(m*K)[41].

While in this case study we focus on calculating z-directional thermal conductivity, any macroscale property that can be computed directly from MD simulations can be substituted into the virtual experiment framework. In a later section, we showcase the flexibility of RAPSIDY 2.0 by providing a second case study focusing on designing high tensile strength block copolymers, in which tensile calculations are performed during virtual experiments as opposed to thermal conductivity calculations as described above.

### D. The Active Learning Loop using Constrained Bayesian Optimization (cBO)

After the morphological stability and physical property is computed for each sample in the initial DoE using virtual experiments, the next step in the process is to initialize and begin the active learning loop. Active learning is a semi-supervised machine learning methodology that "learns" from data to optimally navigate the design space and propose promising candidates for guiding experiments and computations. In our work, we utilize an active learning framework known as constrained Bayesian Optimization (cBO) to intelligently navigate the large polymer design space and to propose new designs to test via virtual experiments. Here, we give a summary of cBO and encourage readers to refer to **Supporting Information section S.I.** and cited references for detailed theoretical derivations on the machine learning methodology.

Traditionally, Bayesian optimization (BO) is a framework that focuses on solving the global optimization problem

$$\max_{\vec{x} \in A} f(\vec{x}) \tag{9}$$

where f(x) is an expensive to evaluate function over the input, x, which belongs to the feasible set, A. The function, f, is typically referred to as a "black box function" as it lacks known structure like concavity or linearity, making traditional optimization techniques, such as derivative-based algorithms, unfeasible. Material properties, such as thermal conductivity as we are studying in this case study, can often be referred to as a "black box function" because the underlying structure-property relationship are unknown. Data is collected and a cheap-to-evaluate surrogate model is trained, which is typically an appropriate Gaussian process (GP) prior of the form:

$$p(f) = GP(f; \mu, K) \tag{10}$$

Observations (X, f) are used the condition the distribution on the data D:

$$p(f|D) = GP(f; \mu_{f|D}, K_{f|D}) \tag{11}$$

Here, $\mu_{f|D}$ refers to the mean of the function conditioned on D and $K_{f|D}$ refers to the covariance of the function conditioned on D. The most common covariance kernel used in BO is the Matern covariance function[42].

Given these observations, the next point in the design space to query is found by evaluating an inexpensive function, known as an acquisition function (AF), which quantifies how desirable evaluating *f* is at a given point for solving our optimization problem. By optimizing the acquisition function, we can select the location within the design space for our next observation. It is important to note that this is an optimization problem on a much cheaper-to-evaluate AF, as compared to our original function, *f*.

However, the focus of our work is to solve the constrained global optimization problem

$$\max_{c(\vec{x}) \leq \lambda} f(\vec{x}) \tag{12}$$

where $c(\vec{x})$ is an expensive to evaluate constraint function and $\lambda$ is the upper bound of the constraint. In our work, $f(\vec{x})$ is the desired material property of interest and our constraint is, $V_r$, the morphological stability with $V_{r,\,cut}$ as the user prescribed cutoff value that decides "stable" or "unstable". In our work, we choose $V_{r,\,cut} = 4.5$. We utilize constrained Bayesian Optimization

(cBO) as opposed to traditional BO as it is much more efficient in navigating the block copolymer design space. BO is inefficient because (1) the phase diagram for block copolymers typically contains narrow regions of stability that are not known *a priori* and (2) macroscopic material properties are often dependent on the nanoscale assembled morphology, which is also not known *a priori*. Thus, BO will simply optimize the given black box objective with no knowledge of the intrinsic stability of the morphology at play.

We follow the constrained optimization framework developed by Gardner *et al.* and utilize the constrained expected improvement (cEI) acquisition function to query the next point in the design space that will be tested via virtual experiment[43]. The details of the derivations of cEI acquisition function can be found in the **Supporting Information section S.I.**. We model both $c(\vec{x})$ and $f(\vec{x})$ as conditionally independent Gaussian processes. In our work, we continue to evaluate the physical property even for design points that are found to be unviable (i.e., $V_r > V_{r,\,cut}$) as they improve the accuracy of the surrogate Gaussian process posteriors. This is especially important in design spaces with small viable regions, which, as previously mentioned, is often the case for many morphologies in phase diagrams for block copolymers[44]. We also highlight that we incorporate expert knowledge to specify possible target morphologies, rather than utilizing cBO to select optimal morphologies because efficiently optimizing acquisition functions with mixed categorical (i.e., morphology) and continuous (i.e., volume fraction) variables is currently an active area of research in the machine learning space[45,46].

III. **Results and Discussion**

A. **Case Study 1: Designing High Thermal Conductivity Block Copolymer Blends**

**Figure 4** shows an example of the active learning loop applied to design high thermal conductivity diblock copolymer blends (design space shown in **Figure 2**) using a target morphology of lamellae (L), double gyroid (DG), and hexagonally packed cylinders (C6). Surrogate models are trained for each target morphology using 100 data points sampled during DoE. We perform independent virtual experiments to determine the morphological stability of the initially sampled points and perform 100 iterations of active learning. One iteration of active learning is defined as a single loop of surrogate model training, acquisition function optimization to determine the next design point, and finally testing via virtual experiment. The data acquired from subsequent virtual experiments is appended to the initial data retrieved from virtual experiments of DoE to improve the accuracy of the surrogate model.

**Figure 4a** shows the average and standard deviation of the z-direction thermal conductivity for the top 10 viable candidates for L (blue), DG (orange) and C6 (green) over the course of 100 active learning iterations. The data point at iteration 0 shows the statistics for the top 10 viable candidates after initial DoE. A viable candidate is a candidate that is found to have met the morphological stability threshold of $V_r < 4.5$, which suggests that the chosen chain design is stable within the target morphology. The data shows that for the design space of BCP blends in **Figure 2,** the values of thermal conductivity in the z-direction show little variation with repsect to the chosen target morphology. We also find that over the course of 100 active learning iterations, the thermal conductivity improves by less than 10%, suggesting that thermal transport properties are a weak function of the chosen design space. **Figure 4b** shows the number of viable candidates found using cBO as a function of the number of active learning iterations. We find that cBO effectively limits the design space search to regions that are most likely to exhibit the targeted morphology, with a ~70%, ~60%, and ~30% success rate for L, DG, and C6, respectively. The lamellar regime of BCP phase diagrams is typically much larger than that of DG and C6 for diblock copolymers and that explains the higher success rate for L.

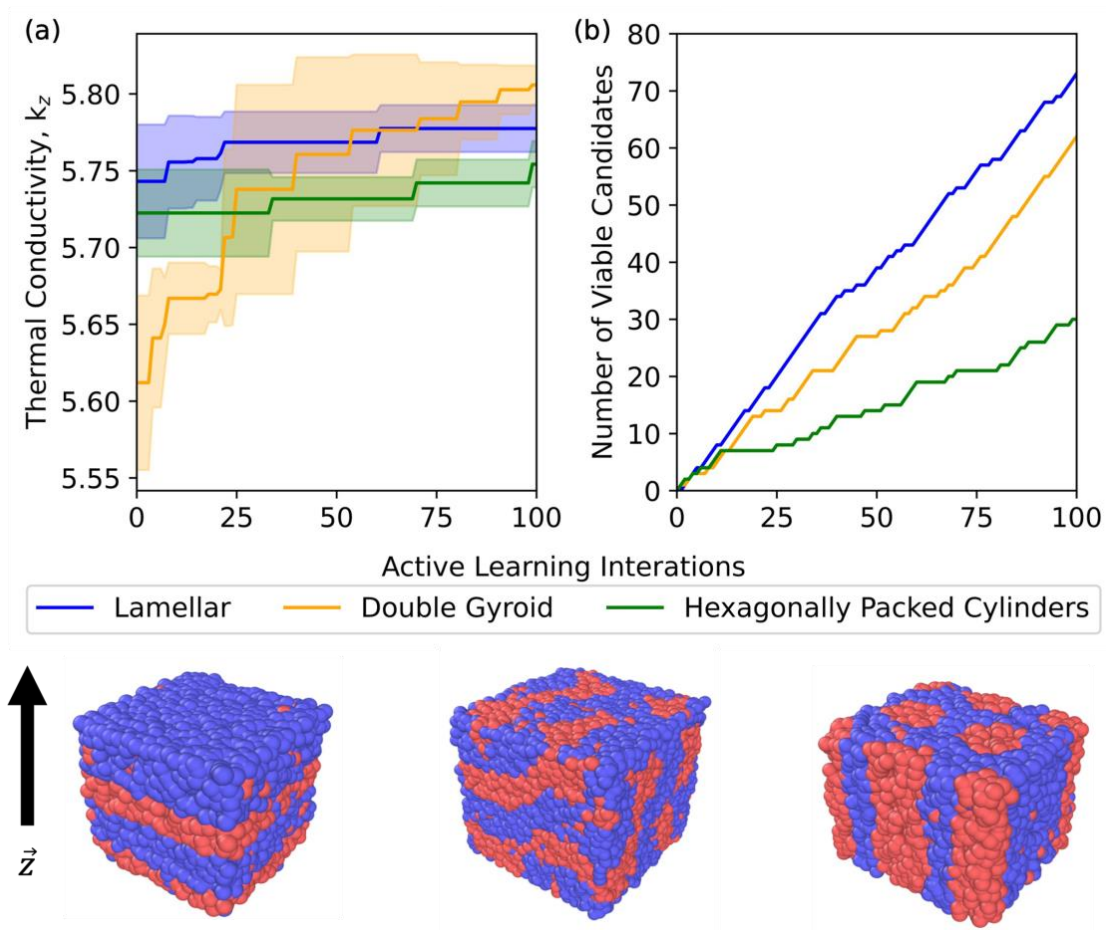

Figure 4. **Results of constrained Bayesian optimization of thermal conductivity for different target morphologies.** (a) Average thermal conductivity and standard deviations for the top 10 candidates for lamellar, double gyroid, and hexagonally packed cylinders after 100 active learning iterations. (b) The number of viable candidates meeting the $V_r$ threshold found using cBO as a function of active learning iterations. cBO effectively limits the search to design parameters that exhibit the target morphology.

Previous studies of thermal conductivity for polymeric systems have shown that the thermal transport properties are highly correlated with chain conformations, specifically the radius-of-gyration ($R_g$). For example, simulations by Wei *et al.* found that amorphous polymers with higher levels of chain extension exhibited higher values of thermal conductivity[47]. Another study by *Zhou et al.* on polyethylene-polypropylene copolymers similarly found that increased chain extension enhanced thermal transport, and that thermal conductivity scaled linearly with average radius-of-gyration[16]. These previous findings motivated us to modify our design space from purely *flexible* diblock copolymer blends (**Figure 2**) to semi-flexible diblock copolymer blends by incorporating block stiffening of species A using a cosine angle potential of the form:

$$U_{angle}(\theta) = K[1 + \cos(\theta)] \tag{13}$$

Here, θ is the angle between two bonded vectors and K is a varying force constant that controls the stiffness of the A blocks. Past studies using coarse-grained simulations on homopolymers have shown how the value of K is related to the persistence length of polymers[48]. We set K as a parameter within our design space and bound K from [0, 20]. For this study, we keep B blocks to be completely flexible (i.e., K=0). The new design space is now shown in **Table 1.**

Table 1. **Design parameter space and upper and lower bounds for semi-flexible diblock copolymer blends with B block being flexible and A block stiffness controlled by the angle potential's force constant K**

| Design Parameter | Bounds |
| --- | --- |
| $N_1$ | [50, 100] |
| $N_2$ | [50, 100] |
| $f_{A1}$ | (0, 1) |
| $f_{A2}$ | (0, 1) |
| $\varphi_1$ | (0, 1) |
| L | [20, 40] |
| K | [0, 20] |

We now repeat our optimization study for semi-flexible diblock copolymer blends, following the same procedure as that of flexible diblock copolymer blends that we presented above. We initially sample 100 samples using DoE in the now 7-dimensional design space (as compared to the 6-dimensional design space previously for the fully flexible copolymer blends) and perform virtual experiments to determine the morphological stability and z-directional thermal conductivity for target morphologies of L, DG, and C6. We provide the sampled values of the design space using DoE for these semi-flexible diblock copolymer blends in the **Supporting Information Figure S1**. For the semi-flexible diblock blends, **Figure 5a** shows the average and standard deviation of the z-direction thermal conductivity for the top 10 viable candidates with target morphologies of L (blue), DG (orange) and C6 (green) over the course of 100 active learning iterations. We now see a 50% increase in z-directional thermal conductivity as compared to the flexible diblock blends (**Figure 4a**) as well as a stratification of $k_z$ due to morphology. We see that lamellar morphology outperforms the curved interface morphologies of C6 and DG by ~10% after 100 active learning iterations.

To better elucidate the effects of incorporating semi-flexibility on $k_z$, we analyzed the parameters sampled through each active learning iteration. **Figure 5b** shows the strength of the angle potential, K, that is sampled over the course of the number of virtual experiments performed with L as the target morphology. The initial 100 virtual experiments belong to the initial DoE data and uniformly samples K ∈ [0, 20] as expected. The final 100 virtual experiments were performed with guidance of the cBO active learning loop and sample values of K close to the upper boundary of K = 20. The active learning loop has "learned" that maximizing the stiffness of the A blocks leads to maximum values of thermal conductivity. We see similar behavior of sampling high K values during active learning for both DG and C6 (**Supporting Information Figure S2 and S3**).

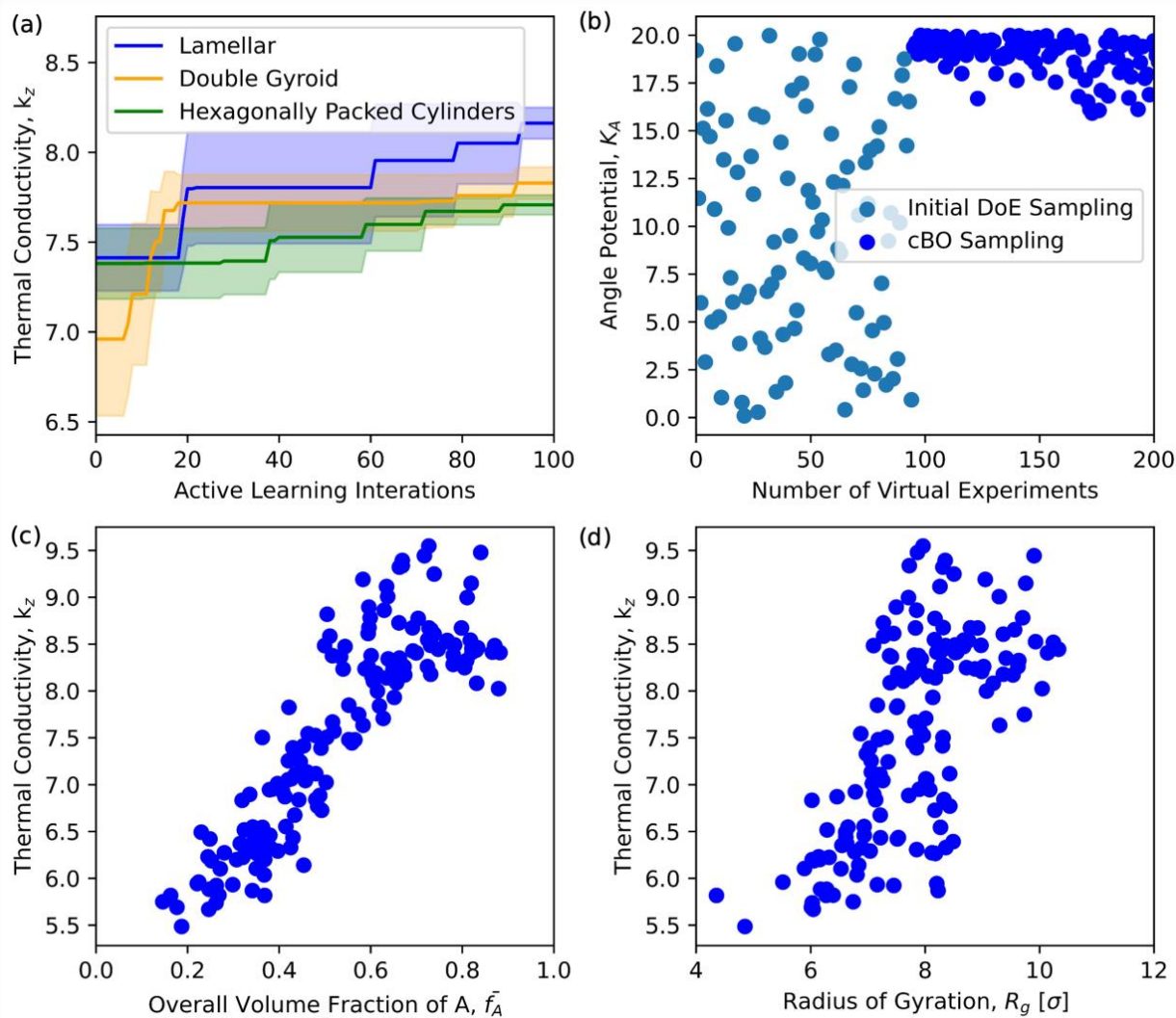

*Figure 5.* **Results of constrained Bayesian optimization for different target morphologies with varying A block flexibility.** (a) Average thermal conductivity and standard deviations for the top 10 candidates for lamellar, double gyroid, and hexagonally packed cylinders after 100 active learning iterations. (b) The value of the sampled angle potential during initial DoE (light blue) and during active learning (blue) for lamellar target morphology. Active learning seeks to maximize chain stiffness to maximize thermal conductivity. (c) Thermal conductivity as a function of overall volume fraction of A during active learning for lamellar

target morphology. Samples that maximize the volume of stiff blocks within the system lead to higher values of thermal conductivity. (d) Thermal conductivity as a function of radius-of-gyration of the entire copolymer chain during active learning for lamellar target morphology.

**Figure 5c** shows $k_z$ as a function of the overall volume fraction of A, $\bar{f_A}$, (which we calculate as $\bar{f_A} = \frac{\varphi_1 N_1 f_{A1} + (1-\varphi_1) N_2 f_{A2}}{\varphi_1 N_1 + (1-\varphi_1) N_2}$) for the designs sampled during the active learning loop for the L target morphology. We see that there is a linear correlation between $\bar{f_A}$ and $k_z$, suggesting that samples that maximize the volume of stiff blocks within the system lead to higher values of thermal conductivity. **Figure 5d** shows $k_z$ as a function of average radius-of-gyration of the copolymer chains for the L target morphology. We observe a linear relationship between $k_z$ and $R_g$ that is consistent with expected extension of polymer chain conformations with stiffening of blocks. We see similar linear relationships between $\bar{f_A}$ and $R_g$ with $k_z$ for both C6 and DG target morphologies (**Supporting Information Figures S2 and S3**).

Overall, results in **Figure 5** demonstrate that samples with (1) maximal bond stiffness and (2) maximal fraction of stiff blocks lead to the highest performance in thermal transport. Previous studies on thermal conductivity within polymeric systems have shown that bonded interactions contribute to a majority of the material's thermal conductivity[47]. While all three morphologies studied (L, DG, and C6) are compatible with high A block stiffness blends (i.e., the limit of K=20), we hypothesize that increasing chain stiffness increases chain alignment *parallel* to the direction of thermal transport (z-direction) to a higher extent in the L morphology than DG and C6. The bottom of Figure 4 shows the orientations of the simulation boxes in which the thermal conductivity is calculated. In the L morphology, the stiffened A blocks orientate perpendicular to the lamellar sheets; this causes the chains to become aligned parallel to the z-direction and the direction of heat transfer. However, DG and C6 both have curved interfaces which may disrupt the alignment of the stiffened A blocks, and the alignment directions are not necessarily parallel to the direction of heat transfer. In the case of C6, the cylindrical interfaces are *exclusively* aligned along the z-axis, causing the chains, which are aligned normal to the A-B interface, to become perpendicular to the direction of heat transfer.

To test this hypothesis that increasing chain stiffness increases chain alignment *parallel* to the direction of thermal transport (z-direction) to a higher extent in the L morphology than DG and C6, we quantify the extent of alignment of the A blocks with the z-direction in each of these morphologies. We calculate the $S_2$ bond order parameter, defined as $S_2 = \frac{3}{2} \langle \cos^2 \theta \rangle - \frac{1}{2}$, where θ is the angle between a bond vector and a user-defined reference vector which in our case is (0,0,1), corresponding to the z-axis. The value of $\frac{3}{2}\cos^2\theta - \frac{1}{2}$ varies between -0.5 (corresponding to the bond vector being perpendicular to the reference vector) to +1 (corresponding to the bond vector being parallel to the reference vector). In **Supporting Information Figure S4a**, we show the probability distribution of $\frac{3}{2}\cos^2\theta - \frac{1}{2}$ for all bonds between A-A species for the top 10 candidates found for L, DG, and C6. We see that C6 has the highest probability density for bond vectors being perpendicular to the z-axis (i.e., $\frac{3}{2}\cos^2\theta - \frac{1}{2} \to -\frac{1}{2}$), followed by DG and L. As expected L has the highest probability density for bond vectors being parallel to the z-axis, the direction of heat transfer (i.e., $\frac{3}{2}\cos^2\theta - \frac{1}{2} \to 1$), followed by DG, and L. The average $S_2$ bond order parameter also ranks from L > DG > C6 with 0.25, 0.16, and 0.13, respectively, showing that a stratification of the extent of preferential orientation (with respect to the z-axis) exists between different morphologies. We also repeated the same $S_2$ calculation as above to bonds between B-B species to determine the effect of flexible blocks on chain orientation (**Figure S4c**). As expected, the distributions are statistically identical, with $S_2$ values essentially equal to 0, suggesting that the B blocks have no preferential orientation. One should note that in the above comparison, the top candidates for L, DG, and C6 may have different values of A block stiffnesses (in other words, different angle potential force constant for A block, K) and composition of A blocks ($\bar{f}_A$).

To compare L, DG, and C6 morphologies' thermal conductivity at constant values of A block stiffnesses, we now perform the thermal conductivity optimization loop on semi-flexible diblock blends of fixed A block stiffness with fixed values of K=0, 10, and 20, which we designate as low, medium, and high stiffness. **Figures 6a and 6b** shows the average and standard deviation of the z-direction thermal conductivity for the top 10 viable candidates for semi-flexible diblock blends with target morphologies of L (blue), DG (orange) and C6 (green) over the course of 100 active learning iterations for high and medium stiffness, respectively. The K=0 case has already been studied in the flexible diblock blend case shown in **Figure 4a**. First, we find that increasing K leads to a monotonic increase in $k_z$, which is consistent with our previous study where K was treated as a varying design parameter. Second, we find that increasing chain stiffness leads to increasing stratification of the optimized thermal conductivity due to chosen target morphology. At low stiffness, the chosen target morphology plays little to no role in determining the thermal conductivity of the material, with differences of $k_z$ between the best and worst performing morphologies being ~0.05. However, increasing the stiffness to medium and high shows a stratification of $k_z$ to 0.4 and 0.6, respectively, between the best (L) and worst (C6) performing morphologies.

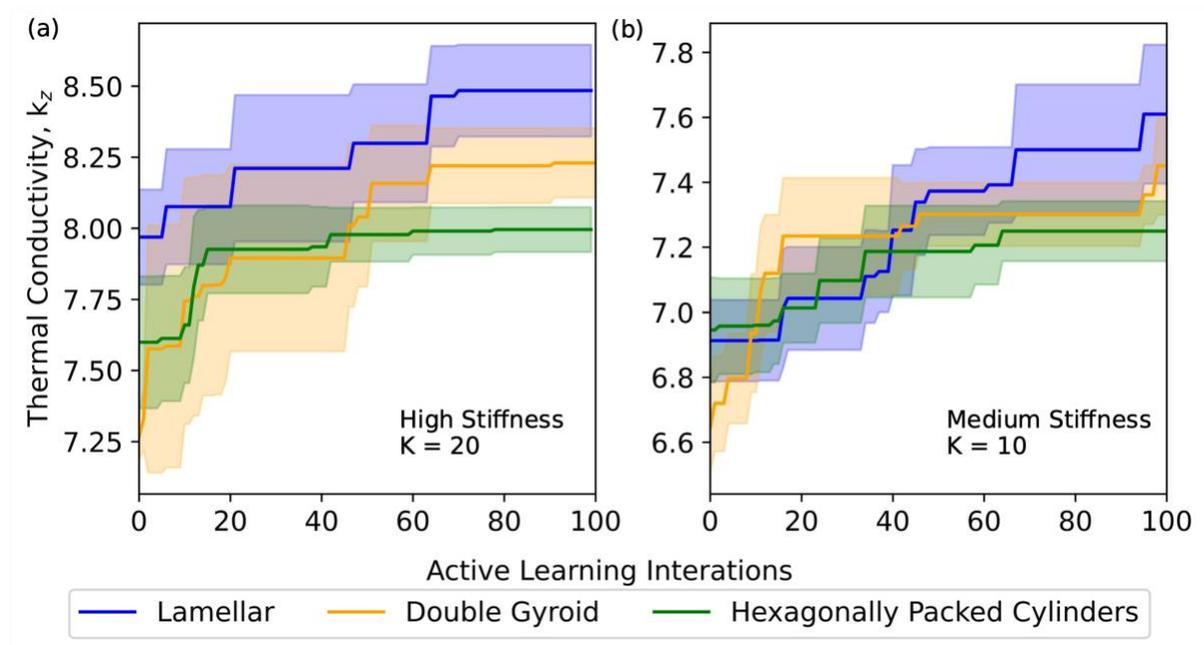

*Figure 6.* **Results of Constrained Bayesian Optimization for different target morphologies with fixed stiffness.** High stiffness, K=20, (a) and medium stiffness, K = 10, (b). Refer to Figure 4 for flexible chains with low stiffness, K = 0.

In conclusion, for this case study focused on thermal conductivity as the target physical property, we find that at the limit of flexible chains the BCP blends morphology plays little to no role in determining thermal transport properties but the type of morphology becomes increasingly important as block or chain stiffness increases. The morphological thermal transport performance hierarchy of lamellar>double gyroid>hexagonally packed cylinders becomes increasingly stratified as chain stiffness increases.

### B. Case Study 2: Designing High Tensile Strength Block Copolymer Blends

Having described the details of our RAPSIDY 2.0 workflow in the previous sections using an example of investigating high thermal conductivity block copolymer blends, we now showcase its efficacy in designing BCPs with superior mechanical properties. Due to their unique ability to microphase separate, BCPs can inherently balance material toughness and strength, which coupled with high precision synthesis techniques for accurate chain design tunability, make them ideal candidates for high-performance applications[49–51]. By fine tuning design parameters such as chain length, chain architecture, and block composition, coupled with its nanoscale morphology, we can design materials with BCPs that exhibit enhanced mechanical properties. In this section, we focus on a case study optimizing the tensile strength of diblock copolymer blends with the same 6-dimensional design space system shown in **Figure 2**.

We follow the same three steps RAPSIDY 2.0 protocol of *initial DoE,* followed by performing *virtual experiments*, and finally utilizing the *active learning loop* to optimally query the large design space. However, we now modify our *virtual experiments* from calculating thermal conductivity using NEMD to performing uniaxial deformation simulations. Uniaxial tension simulations deform the simulation box from $L_0$ to $L_0 a$ in the direction of tensile deformation and from $L_0$ to $L_0 a^{-1/2}$ in the orthogonal directions at a constant rate (here, $L_0$ is the original box length and a is the tensile elongation) to preserve the total volume of the system[52]. The average stress in the tensile direction, σ, can be computed using the deviatoric part of the stress tensor. The Young's modulus can be fitted to the linear regime of the resulting stress-strain curve where the material experiences reversible elastic deformation. We developed a high-throughput algorithm to automatically determine the bounds of the elastic deformation region and perform linear fits to determine the resulting Young's modulus. Detailed derivations on the algorithm and an example of a resultant stress-strain curve are shown in the **Supporting Information, section S.III.**. We also run all MD simulations within the *virtual experiment* at a reduced temperature of T*=0.3 such that we are below the polymer's glass transition temperature to allow for plastic stress-strain behavior (as opposed to elastomeric stress-strain behavior).[53]

**Figure 7a** shows the average and standard deviation of the z-direction modulus, $E_z$, for the top 10 viable candidates for diblock copolymer blends with target morphologies of L (blue), DG (orange) and C6 (green) over the course of 100 active learning iterations. We see significant stratification in morphological performance, with C6 shows the highest tensile strength, followed by DG, and L. While C6 shows a significant tensile improvement of 20% over 100 active learning iterations over the initial DoE sampling, DG and L show only ~10% improvement, suggesting that the Young's modulus for DG and L is less sensitive to chain design versus that of C6. The hierarchy of modulus as a function of morphology (C6 > DG > L) is expected and consistent with experimental literature, which shows significant tensile anisotropy for C6 and L morphologies due to their two-dimensional periodicity. For example, Honeker and Thomas synthesized ABA triblock copolymers with the C6 morphology and observed that tensile strength was higher in the direction parallel versus perpendicular to the cylinder axes[54]. On the other hand, Cohen *et al.* synthesized AVA triblock copolymers with the lamellar morphology and observed that tensile strength also was higher in the direction parallel versus perpendicular to the lamellar sheets. In **Figure 7** we show the orientation of the simulation cell; the z-axis deformation we perform is parallel to the cylindrical axes in C6 and thus, exhibits the highest tensile strength. The z-axis deformation is perpendicular to the lamellar sheets in L and thus, exhibits the lowest tensile strength. As a result, we expect significant anisotropic behavior in the tensile strength of C6 and L, but not for DG due to its three-dimensional symmetry. However, as RAPSIDY 2.0 only focuses on single objective optimization, we simply focus on tensile strength in a single direction; multi-objective optimization is the scope of our future work. Finally, **Figure 7b** shows the number of viable candidates found using cBO; our methodology effectively screens for viable L candidates more often compared to C6 and DG due to the large L stability window in the phase diagram of diblock copolymers.

In summary, we showcase the ability of RAPSIDY 2.0 to effectively tackle the inverse problem of designing high-performance BCPs. Our results of designing high tensile strength BCP blends showcase a morphological hierarchy of C6>DG>L and are consistent with trends found in experimental studies. Our methodology can be flexibly adapted to any property of interest, with the condition that it can be computed using MD simulations via simple modification of the in-silico *virtual experiment* procedure.

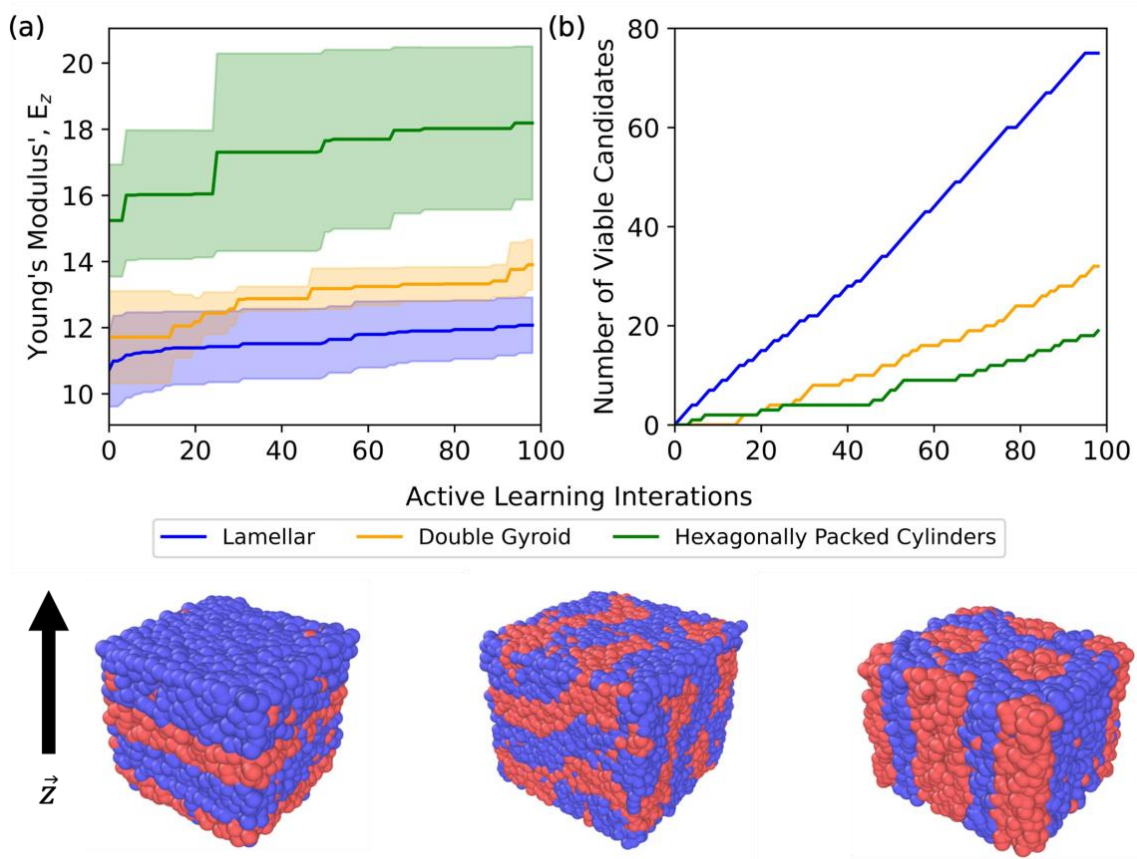

Figure 7. **Results of Constrained Bayesian Optimization of Young's Modulus for different target morphologies.** (a) Average Young's Modulus and standard deviations for the top 10 candidates for lamellar, double gyroid, and hexagonally packed cylinders after 100 active learning iterations. (b) The number of viable candidates meeting the $V_r$ threshold found using cBO as a function of active learning iterations. cBO effectively limits the search to design parameters that exhibit the target morphology.

## IV. Conclusions

In this work, we present an in-silico high throughput inverse design methodology, RAPSIDY 2.0 - Rapid Analysis of Polymer Structure and Inverse Design strategy 2.0, that combines molecular dynamics simulations with Bayesian Optimization driven active learning to design high-performance copolymers. We showcase the efficacy of our workflow on two case studies by identifying designs of blends of diblock copolymers with (1) high thermal conductivity and (2) high tensile strength.

For the first case study on blends of diblock copolymers with high thermal conductivity, we demonstrated the efficacy of our active learning framework to effectively explore the large, high-dimensional parameter space of block copolymers and identify parameters in the design space that impact thermal conductivity and optimal designs of the blends. We utilize molecular dynamics simulations inspired by our previous work, RAPSIDY, to rapidly assess the stability of a targeted nanoscale morphology (lamellae, double gyroid, hexagonally packed cylinders) and compute the material's physical property via virtual (in-silico) experiments. Using a machine learning framework driven by constrained Bayesian Optimization, we demonstrate the efficacy of active learning in proposing promising candidates that simultaneously maximize the target macroscale property and stabilize the targeted morphology for guiding optimal experiments. We find that increasing one of the blocks' stiffness plays a key role in increasing thermal transport. We also find that in the limit of fully flexible BCP chains, the blend morphology plays little to no role in determining thermal transport properties, but blend morphology becomes increasingly important as the chain or one of the blocks' stiffness increases. The morphological thermal transport performance hierarchy of L > DG > C6 becomes increasingly stratified as chain stiffness increases.

For the second case study on designing blends of flexible diblock copolymers with high tensile strength, we demonstrated that morphology plays a key role in determining the materials modulus. We find that there is a hierarchy of directional modulus as a function of morphology (C6>DG>L), which is consistent with findings from experimental literature, which shows significant tensile anisotropy for C6 and L morphologies due to their two-dimensional periodicity. We showcase that RAPSIDY 2.0 can be flexibility implemented for any parameter of interest, with the condition that it can be computed directly from molecular dynamics simulations.

RAPSIDY 2.0 is computationally efficient as it limits the high-dimensional polymer design space search to promising candidates that exhibit both morphology stability and high material performance and can be flexibly adapted to any property of interest, with the condition that it can be computed using MD simulations. RAPSIDY 2.0 is one of the first methodologies that directly addresses the unique challenge associated with designing high-performance copolymers in that their material properties are a function of both their sequence design and nanoscale morphology, which are coupled. We expect our methodology to be highly valuable in guiding high-throughput experimental design and synthesis of the next generation of novel polymeric materials.

**Conflicts of Interest**

The authors declare no competing interests.

**Acknowledgements**

This work was financially supported by the Army Research Office through the Multi-disciplinary University Research Initiative (MURI) award number W911NF2310260. This research was supported in part through the use of Information Technologies (IT) resources at the University of Delaware, specifically the high-performance computing resources.

**Data and Code Availability**

All data needed to evaluate the conclusions in the paper are available either in the main text or Supplementary Information. RAPSIDY 2.0 and example code is available at our GitHub repository, https://github.com/arthijayaraman-lab.

**For Table of Contents Only**

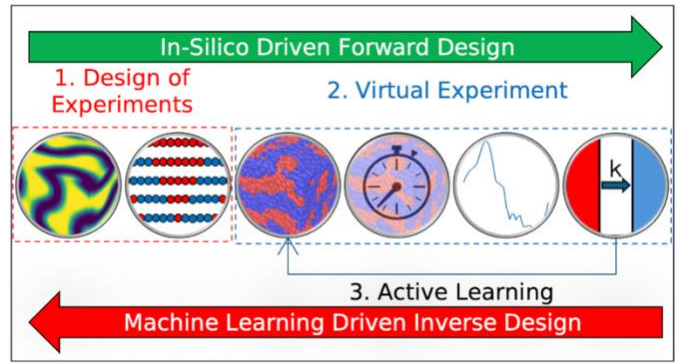

In RAPSIDY 2.0 we use a combination of molecular dynamics simulations and Bayesian optimization driven active learning to optimally query a high-dimensional polymer design space and propose optimal design candidates that simultaneously stabilize a selected nanoscale morphology and exhibit desired macroscale material properties (e.g., tensile strength, thermal conductivity).